\documentstyle[fullpage,epsf,11pt]{article}

\begin{document}
\begin{center}
  {\bf Experimental Demonstration of Greenberger-Horne-Zeilinger
  Correlations \\ Using Nuclear Magnetic Resonance \\}
  \bigskip
    Richard J. Nelson$^\dagger$ \\ David G. Cory$^*$ \\ Seth Lloyd$^\dagger$ \\
    \smallskip
    $\dagger$ d'Arbeloff Laboratory for Information Systems and Technology \\ 
	Department of Mechanical Engineering \\
    $*$ Department of Nuclear Engineering \\
    Massachusetts Institute of Technology \\ Cambridge, Mass. 02139 \\
\end{center}

\bigskip

\noindent{\it Abstract:} 
The Greenberger-Horne-Zeilinger (GHZ) effect provides an example of 
  quantum correlations that cannot be explained by classical local hidden 
  variables.
This paper reports on the experimental realization of GHZ correlations using
  nuclear magnetic resonance (NMR).
The NMR experiment differs from the originally proposed GHZ experiment in
  several ways: it is performed on mixed states rather than pure states; and
  instead of being widely separated, the spins on which it is performed are
  all located in the same molecule.
As a result, the NMR version of the GHZ experiment cannot entirely rule out
  classical local hidden variables.  
It nonetheless provides an unambiguous demonstration of the 
  ``paradoxical" GHZ correlations, and shows that any classical hidden 
  variables must communicate by non-standard and previously undetected 
  forces.
The NMR demonstration of GHZ correlations shows the power of 
  NMR quantum information processing techniques for demonstrating 
  fundamental effects in quantum mechanics.

\bigskip

Quantum mechanics is well-known to exhibit strange and apparently
  paradoxical effects.
Perhaps the strangest of these effects is the possibility of correlations
  between quantum-mechanical systems that cannot be explained by appealing
  to an underlying theory of local hidden variables---classical variables
  whose unknown or ``hidden'' states supply the apparent stochastic results to
  quantum experiments
  \cite{epr:cqmdrcc,b:qt,b:asiqtthv,b:oeprp,agr:etrltvbt,agr:ereprbg}.
A striking contradiction with the local hidden variable picture is
  obtained in the Greenberger-Horne-Zeilinger (GHZ) effect, in which a set of 
  measurements on three correlated quantum particles gives results that are
  perfectly correlated in a pattern that is classically impossible
  \cite{ghsz:btwi}.
This paper reports the first experimental demonstration of GHZ correlations.

GHZ experiments that verify both the ``weird" quantum correlations and the
  non-local nature of physical reality are difficult to realize in the
  laboratory, where the problems associated with faithfully preparing the
  required state $\vert\psi\rangle$ and making the necessary measurements pose
  formidable obstacles.
However, recent discoveries in the field of quantum information processing
  using NMR make possible an analog GHZ experiment that validates the GHZ
  correlations required by a quantum mechanical description of the universe
  \cite{gc:bsrqc,cfh:eqcnmrs,cph:nmrsaeapqc}.
The idea of using NMR to demonstrate GHZ correlations was proposed in
  \cite{l:maghze}, and an effective GHZ state was first created using NMR
  in \cite{lkzcm:nmrghz}.
NMR offers a number of advantages in performing tests of few-particle
  quantum mechanics, including:
  (1) weak measurements that allow the direct determination of ensemble
    density matrices;
  (2) long decoherence times;
  (3) simple, well-understood Hamiltonians; 
  (4) gradient techniques that allow the construction of effective
    pure states and the controlled decoherence of samples.
These features of NMR make it possible to use quantum systems in mixed states
  at room temperature to mimic precisely the dynamics of a quantum system in
  a pure state at low temperature.  
As a result, NMR is an ideal testbed for the predictions of quantum mechanics
  for few-variable systems.
 
Despite the advantages of NMR quantum information processing techniques
  listed above, it is important to note that demonstrating GHZ correlations
  using NMR differs in significant ways from performing the GHZ experiment
  as originally envisaged.
In particular, because the NMR experiment is inherently local, the NMR
  demonstration of GHZ correlations cannot rule out nonstandard interactions
  between classical hidden variables that conspire to exhibit apparently
  quantum mechanical correlations.
In addition, because the NMR experiment is performed on thermal states,
  though the spins in the experiment mimic the effects of entanglement, they
  are not in fact entangled: as a result, the NMR experiment cannot rule out
  local hidden variables.
Nonetheless, the NMR demonstration of GHZ correlations provides an unambiguous
  and statistically significant test of the predictions of quantum mechanics.
The experiment shows these to be accurate to a high degree of precision.  

The following version of the GHZ experiment is due to Mermin \cite{m:qmr}.
Consider three spin-$1/2$ particles prepared in the state
 \begin{equation}
   \vert\psi\rangle=\frac{1}{\sqrt{2}}\left(\vert{\rm +++}\rangle\
                                        -\ \vert{\rm ---}\rangle\right)\ ,
   \label{eq:GHZ}
 \end{equation}
where $\vert {\rm +}\rangle$ represents ``up" along the $z$-axis and
  $\vert {\rm -}\rangle$ signifies ``down" along the $z$-axis.
Now consider the results of the following products of spin measurements, each
  made on state $\vert\psi\rangle$:

\bigskip

(i) particle (1) along $x$, particle (2) along $y$, and particle (3) along $y$.
Note that since
  $\sigma_x^1\sigma_y^2\sigma_y^3\vert\psi\rangle=+\vert\psi\rangle$,
  the product of the results of these measurements should be $+1$, where
  $+1$ corresponds to the result $\vert {\rm +}\rangle$, and $-1$ corresponds
  to $\vert {\rm -}\rangle$.

(ii) particle (1) along $y$, particle (2) along $x$, and particle (3) along
  $y$.
As in scenario (i), since
  $\sigma_y^1\sigma_x^2\sigma_y^3\vert\psi\rangle=+\vert\psi\rangle$, the
  product of these results should be $+1$.

(iii) particle (1) along $y$, particle (2) along $y$, and particle (3) along
  $x$.
Again, since
  $\sigma_y^1\sigma_y^2\sigma_x^3\vert\psi\rangle=+\vert\psi\rangle$,
  the product of these results should also be $+1$.

(iv) particle (1) along $x$, particle (2) along $x$, and particle (3) along
  $x$.
In this case, there is a significant sign difference:
  $\sigma_x^1\sigma_x^2\sigma_x^3\vert\psi\rangle=-\vert\psi\rangle$, so the
  product of the results should be $-1$.

\bigskip

The minus sign in the final scenario is crucial for 
  differentiating between quantum mechanical and classical ({\it i.e.}
  hidden-variable) descriptions of reality.
No classical hidden-variable model that assigns a classical variable
  (such as $+1$ or $-1$) to the measurement of each individual particle's
  spin along axes $x$ and $y$ could reproduce the quantum predictions:
  classically, the product of the four scenarios above for a hidden-variable
  model must yield $+1$ because each particle is measured along each of the
  two axes exactly twice.

In addition to demonstrating purely quantum correlations that have no classical
  analog, the above set of measurements has the potential to verify the
  hallmark of quantum ``weirdness": non-locality.
The GHZ experiment---as originally conceived---requires that the measuring
  apparatuses for the three particles be sufficiently distant from each
  other so as to eliminate the possibility of local, classical interactions
  between measurements that might duplicate the quantum mechanical predictions.
In NMR experiments, the effective distances between any three spins in a
  molecule are typically a few angstroms at most; as such, it is practically
  impossible to demonstrate non-local effects using NMR.
However, in the NMR experiment below, the system's Hamiltonian is well known
  to a high degree of accuracy; thus, the NMR experiment can rule out any
  \/{\em standard} local interaction as the cause of the non-classical
  correlations.

This paper reports on the first experimental demonstration of GHZ correlations
  using effective pure states.
Techniques now exist whereby it is possible to create an {\em effective} pure
  state in a liquid sample of ${\cal O}(N_A)$ essentially non-interacting
  molecules at room temperature, meaning that the state of the ensemble can be
  described by a density matrix whose deviation from unity transforms under
  unitary operations exactly as a pure state density matrix.
Since only the deviation density matrix ($\rho$), averaged over the whole
  sample, contains observable magnetization, quantum mechanics predicts
  that the results obtained from performing unitary
  transformations on a pseudo-pure NMR state are identical to those that would
  be found were one to perform the same operations on a single quantum system.
For example, the same unitary operator ({\em e.g.},
  ${\cal U}=e^{i\frac{\pi}{4}\sigma_x^1\sigma_x^2\sigma_y^3}$)
  that transforms a pure state ({\em e.g.}, $\vert {\rm +++}\rangle$) into a
  GHZ state ({\em e.g.}, $\vert\psi\rangle$) also transforms the deviation
  density matrix of an ensemble pseudo-pure state into a density matrix with
  the identical, non-classical correlations.

It is important to keep in mind that although the density matrix for an
  ensemble in an effective pure GHZ state is the same as the density matrix for
  an ensemble in which some fraction of the systems are in a pure GHZ state and
  the remainder are in a completely mixed state, these two ensembles are not
  the same.
Two different ensembles may have the same density matrix.
In the ensemble constructed here, each of the systems is in a
  unitarily-transformed version of a thermal state; none are in GHZ states.
Nevertheless, since two ensembles with the same density matrix are
  indistinguishable with respect to measurement, the effective GHZ state
  constructed is just as effective as an actual GHZ state for testing the
  predictions of quantum mechanics.

Our verification of GHZ correlations was performed on a Bruker AMX400
  spectrometer at room temperature.
At this temperature, the Boltzmann distribution over an ensemble of homonuclear
  spins in a liquid NMR sample gives the average density matrix (per molecule)
  approximately proportional to
  ${\bf 1}+\sum_j\beta_j\sigma_z^j$, where
  $\beta_j=2\mu_j{\cal B}/k_BT=h\omega_j/k_BT\approx h\omega/k_BT\approx\beta$
  is the Boltzmann factor for the $j$'th spin.
With $\omega\approx 100$\,MHz for $^{13}{\rm C}$ in a static magnetic field of
  ${\cal B}=9.6$~Tesla, the room-temperature ($T\approx 300$\,K) Boltzmann
  factor is of order $10^{-6}$.
Higher order terms in the density matrix expansion are at least of order
  $10^{-12}$, and can therefore be neglected.
Only the deviation of the density matrix from unity, rescaled in the rest of
  this paper  (at equilibrium, $\rho_{eq}=\sum\sigma_z^j$),
  represents surplus or deficit populations in energy levels whose transitions
  are observable in NMR.

In our experimental setup, one set of coils was available to provide a gradient
  in ${\cal B}_z$ across the sample.
The result of a gradient is to introduce phase variations in the
  sample that vary spatially, which, when averaged over the whole sample,
  represent non-unitary operations on the ensemble density matrix.
In addition, another set of coils in the $x-y$ plane can provide rf pulses that
  cause nearly unitary rotations of the density matrix.
The rf coils are also used as pick-up coils to measure magnetization in the
  $x-y$ plane.
For three spins (as in the demonstration below), the observable magnetization
  is the spatially-averaged signal during NMR signal acquisition that is
  proportional to
  $tr\left[\rho\sum\left(\sigma_x^j+i\sigma_y^j\right)\right]\ .$
Other elements of the density matrix are unobserved.
Correlations between spins can be read directly from the spectra produced by
  plotting the Fourier transform of the induction signal.

The sample used in the demonstration of GHZ correlations was triply-labled
  ($^{13}{\rm C}$) alanine with $^1 {\rm H}$ decoupling.
The lowest-order terms in the natural Hamiltonian for the system are
  \begin{equation}
   {\cal H}\approx\frac{1}{2}\omega_1\sigma_z^1
                 +\frac{1}{2}\omega_2\sigma_z^2
                 +\frac{1}{2}\omega_3\sigma_z^3
                 +\frac{\pi}{2}J_{12}\sigma_z^1\sigma_z^2
                 +\frac{\pi}{2}J_{23}\sigma_z^2\sigma_z^3
  \label{eq:Hamiltonian}
  \end{equation}
  where $\omega_j$ is the Larmor precession frequency (including the effects of
  chemical shift) for spin $j$, and where $J_{jk}$ is the usual scalar (weak)
  coupling between spins $j$ and $k$.
The strengths of the coupling interactions between the carbon spins in
  alanine are $J_{12}=53.4$\,Hz and $J_{23}=35.3$\,Hz.
The scalar interaction between spins $1$ and $3$ is small ($J_{13}=1.4$\,Hz);
  this term and all other interactions in the Hamiltonian (including the
  effects of diffusion and relaxation) can be neglected over the
  $\approx 160$\,ms during which the entire experiment takes place.

The experiment works as follows:

\bigskip

(A) {\em Equilibration}.
The sample is allowed to come to thermal equilibrium, in which the density
  matrix is:
  \begin{equation}
   \rho_{eq}=\sigma_z^1+\sigma_z^2+\sigma_z^3
            =\left(
             \begin{array}{rrrrrrrr}
              3&0&0& 0&0& 0& 0& 0\\
              0&1&0& 0&0& 0& 0& 0\\
              0&0&1& 0&0& 0& 0& 0\\
              0&0&0&-1&0& 0& 0& 0\\
              0&0&0& 0&1& 0& 0& 0\\
              0&0&0& 0&0&-1& 0& 0\\
              0&0&0& 0&0& 0&-1& 0\\
              0&0&0& 0&0& 0& 0&-3
             \end{array}\right)\ .
   \end{equation}

(B) {\em Preparation}.
The sample is prepared in an incoherent mixture of two pseudopure states using
  the following operations:
  \begin{eqnarray*}
   \rho_{eq}\longrightarrow
   \left[\frac{\pi}{2}\right]_{90^{\circ}}^2
   &-&\left[\frac{\partial {\cal B}_z}{\partial z}\right]\\
   &-&\left[\frac{\pi}{2}\right]_{-90^{\circ}}^3-
   \left[\frac{1}{4J_{23}}\right]-
   \left[\frac{\pi}{2}\right]_{135^{\circ}}^3-
   \left[\frac{\partial {\cal B}_z}{\partial y}\right]\\
   &-&\left[\frac{\pi}{2}\right]_{-90^{\circ}}^1-
   \left[\frac{1}{4J_{12}}\right]-
   \left[\frac{\pi}{2}\right]_{135^{\circ}}^1-
   \left[\frac{\partial {\cal B}_z}{\partial x}\right]\\
   &-&\left[\frac{\pi}{2}\right]_{0^{\circ}}^2-
   \left[\frac{1}{2J_{12}}\right]-
   \left[\frac{\pi}{2}\right]_{-90^{\circ}}^2\longrightarrow
   \rho_{pp}.
  \end{eqnarray*}
In this notation, each bracketed expression represents one matrix operation.
Brackets containing angles indicate an rf pulse that selectively rotates the
  superscripted spin by the specified angle, where the subscripted ``phase"
  designates the axis of rotation in the $x-y$ plane of the co-rotating frame.
Selective pulses in this experiment required a pulse length of 2\,ms.
Brackets with partial derivatives indicate an applied gradient in the
  magnetic field along the direction in the denominator.
Brackets containing a coupling constant indicate free evolution of the system
  under the natural Hamiltonian for the time indicated; this evolution allows
  the pertinent scalar coupling term to correlate spins.
Pulses for refocussing or for averaging to zero other coupling terms in the
  natural Hamiltonian are not shown.

The result of the preparation sequence is the density matrix:
  \begin{equation}
   \rho_{pp}=\left(
             \begin{array}{rrrrrrrr}
              1&0&0&0&0&0&0& 0\\
              0&0&0&0&0&0&0& 0\\
              0&0&0&0&0&0&0& 0\\
              0&0&0&0&0&0&0& 0\\
              0&0&0&0&0&0&0& 0\\
              0&0&0&0&0&0&0& 0\\
              0&0&0&0&0&0&0& 0\\
              0&0&0&0&0&0&0&-1
             \end{array}\right)\ ,
  \end{equation}
  which transforms like a balanced mixture of two pseudopure states,
  \begin{equation}
    \frac{1}{2}\left(\vert {\rm +++}\rangle\langle {\rm +++}\vert\ -\ 
                     \vert {\rm ---}\rangle\langle {\rm ---}\vert\right)\ .
  \end{equation}

(C) {\em Rotation}.
The density matrix for the sample is now rotated via unitary operator
  $e^{i\frac{\pi}{4}\sigma_x^1\sigma_x^2\sigma_y^3}$ using the sequence:
  \begin{displaymath}
   \rho_{pp}\longrightarrow
   \left[\frac{\pi}{2}\right]_{-90^{\circ}}^1-
   \left[\frac{1}{2J_{12}}\right]-
   \left[\frac{\pi}{2}\right]_{0^{\circ}}^{2,3}-
   \left[\frac{1}{2J_{23}}\right]-
   \left[\frac{\pi}{2}\right]_{180^{\circ}}^{2,3}-
   \left[\frac{1}{2J_{12}}\right]-
   \left[\frac{\pi}{2}\right]_{-90^{\circ}}^1\longrightarrow
   \rho_{GHZ}\ .
  \end{displaymath}
The density matrix thus rotated becomes
  \begin{equation}
   \rho_{GHZ}=\left(
              \begin{array}{rrrrrrrr}
                0&0&0&0&0&0&0&-1\\
                0&0&0&0&0&0&0& 0\\
                0&0&0&0&0&0&0& 0\\
                0&0&0&0&0&0&0& 0\\
                0&0&0&0&0&0&0& 0\\
                0&0&0&0&0&0&0& 0\\
                0&0&0&0&0&0&0& 0\\
               -1&0&0&0&0&0&0& 0
              \end{array}\right)\ .
  \end{equation}
Note that this matrix has the identical off-diagonal structure as the density
  matrix formed from the pure state of Eq.~\ref{eq:GHZ}:
  \begin{equation}
    \rho_{GHZ}=\frac{1}{2}\left(-\ \vert{\rm +++}\rangle\langle{\rm ---}\vert\ 
                                -\ \vert{\rm ---}\rangle\langle{\rm +++}\vert
                          \right)\ .
    \label{eq:ppGHZ}
  \end{equation}
  {\em i.e.}, the NMR sample as prepared above represents the same quantum
  correlations that would be present in a pure, three-particle GHZ state.

(D) {\em Measurement}.
An NMR measurement that demonstrates one of the
  $\left(\sigma_x^1\sigma_y^2\sigma_y^3\right)$,
  $\left(\sigma_y^1\sigma_x^2\sigma_y^3\right)$,
  $\left(\sigma_y^1\sigma_y^2\sigma_x^3\right)$, or
  $\left(\sigma_x^1\sigma_x^2\sigma_x^3\right)$ correlations is now performed
  by applying rf pulses to the sample to rotate the desired correlation into
  observable magnetization.
Specifically, to measure $\left(\sigma_j^1\sigma_k^2\sigma_l^3\right)$, the
  pulse sequence
  \begin{displaymath}
    \rho_{GHZ}\longrightarrow
    \left[\frac{\pi}{2}\right]_{j-90^{\circ}}^1-
    \left[\frac{\pi}{2}\right]_{l-90^{\circ}}^3
  \end{displaymath}
  followed by data acquisition with the absorptive signal phased along $k$ will
  reveal the magnitude and sign of the desired correlation in the spectrum on
  resonance with the second spin.
Steps (A)$\sim$(C) of the experiment are then identically repeated three times,
  each with a different measurement sequence (D).

\bigskip

The results of the four measurements on the sample prepared in state
  $\rho_{GHZ}$ are shown in Fig.\,1.
For example, to determine the correlations between
  $\left(\sigma_x^1\sigma_y^2\sigma_y^3\right)$, look at Fig.\,1(a):
  the spectrum shows a multiplet of four lines.
The frequency of each of the four lines corresponds to the resonant frequency
  of the second spin when the first and third spins are, respectively (from
  left to right):
  $\vert -\rangle_1\vert -\rangle_3$,
  $\vert +\rangle_1\vert -\rangle_3$,
  $\vert -\rangle_1\vert +\rangle_3$, and
  $\vert +\rangle_1\vert +\rangle_3$.
The fact that the phase of the first line from the left is (+) confirms
  the GHZ correlation that when spin 1 is $\vert -\rangle$ along $x$ and spin
  3 is $\vert -\rangle$ along $y$, spin 2 is $\vert +\rangle$ along $y$.
Similarly, the second line (from the left) shows a negative peak, confirming 
 the correlation that when spin 1 is $\vert +\rangle$ along $x$ and spin 3 is
  $\vert -\rangle$ along $y$, spin 2 is $\vert -\rangle$ along $y$.
Likewise, the third line, with a negative peak, confirms the GHZ correlation
  that when spin 1 is $\vert -\rangle$ along $x$ and spin 3 is
  $\vert +\rangle$ along $y$, spin 2 is $\vert -\rangle$ along $y$.
Finally, the last line confirms the correlation that when spin 1 is
  $\vert +\rangle$ along $x$ and spin 3 is $\vert +\rangle$ along $y$, spin
  2 is $\vert +\rangle$ along $y$.
Thus all four lines in Fig.\,1(a) confirm that the product of the spin of
  particle 1 along $x$, particle 2 along $y$, and particle 3 along $y$
  yields the result $+1$, as in scenario (i) of the GHZ experiment
  explained above.

\begin{figure}[htp]
  \begin{center}
    \epsfxsize=1\hsize
    \leavevmode\epsffile{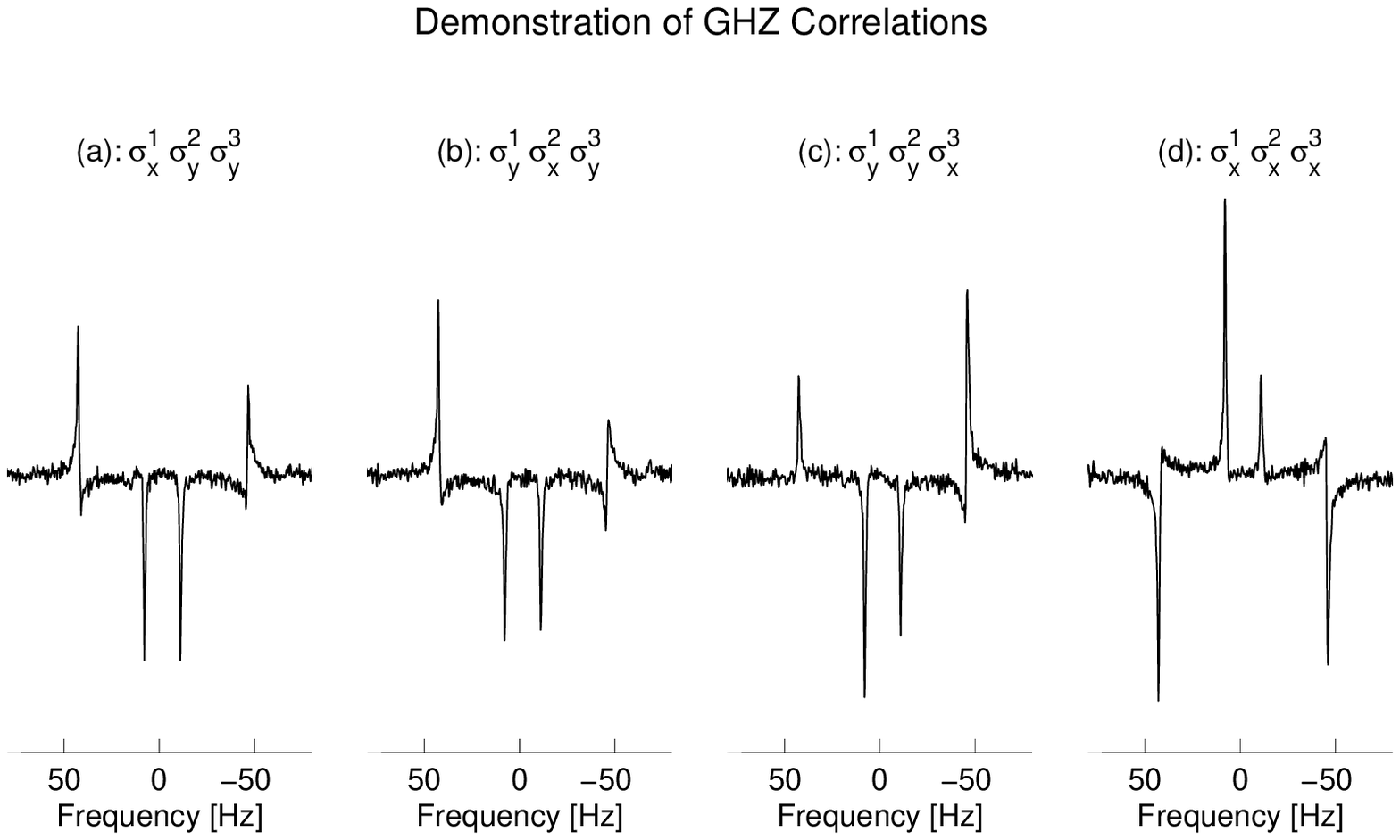}
  \end{center}
  \caption{The confirmation of the quantum prediction of GHZ correlations.
    The four figures show the spectrum of spin 2 for each of the measurements.
    Each spectrum is split into a multiplet of four lines, corresponding
      to the resonant frequencies for spin 2 when spins 1 and 3 are in
      the states $\vert-\rangle_1\vert-\rangle_3$,
                 $\vert+\rangle_1\vert-\rangle_3$,
                 $\vert-\rangle_1\vert+\rangle_3$, and
                 $\vert+\rangle_1\vert+\rangle_3$ respectively.
    An `up' line indicates the state $\vert+\rangle$ for spin 2, and a `down'
      line indicates the state $\vert-\rangle$.
    Plot (a) confirms the GHZ correlation that the product of measurements
      of $\sigma_x$ on the first spin, $\sigma_y$ on the second spin, and
      $\sigma_y$ on the third spin yields the result $+1$.
    For example, the first line in (a) indicates that when spins 1 and 3
      give the result $(-1)_1(-1)_3$, spin 2 gives the result $+1$; the
      second line indicates that when spins 1 and 3 give the result
      $(+1)_1(-1)_3$, spin 2 gives the result $-1$; {\it etc.}, so that the
      product of the results is always $+1$.
    Similarly, (b) and (c) confirm that the products
      $\sigma^1_y\sigma^2_x\sigma^3_y$ and $\sigma^1_y\sigma^2_y\sigma^3_x$
      also yield the result $+1$.
    Plot (d), by contrast, shows that the product
      $\sigma^1_x\sigma^2_x\sigma^3_x$ yields the result $-1$, in
      contradiction to the predictions of the classical hidden variable
      theory.}
\end{figure}

Note that Fig.\,1(b) shows the same four lines in the spectrum for spin
  2, except that in this case the measurement sequence reveals the correlation
  that corresponds to $\sigma_y^1\sigma_x^2\sigma_y^3$.
As in Fig.\,1(a), all four lines indicate that the product of the spin of
  particle 1 along $y$, particle 2 along $x$, and particle 3 along $y$ is
  $+1$, as predicted in scenario (ii) above.
Likewise, Fig.\,1(c) shows the analogous $+1$ result when the product of
  spin measurements of particle 1 along $y$, particle 2 along $y$, and
  particle 3 along $x$ is computed.

Significantly, Fig.\,1(d) shows a different spectrum when a measurement of
  the correlation $\left(\sigma_x^1\sigma_x^2\sigma_x^2\right)$ was made.
In this case, when spin 1 is $\vert -\rangle$ along $x$ and spin 3 is
  $\vert -\rangle$ along $x$, spin 2 is $\vert -\rangle$ along $x$, as seen
  in the first line; when spin 1 is $\vert +\rangle$ along $x$ and spin 3 is
  $\vert -\rangle$ along $x$, spin 2 is $\vert +\rangle$ along $x$; and so
  forth.
In this case, the product of the spin measurements of particle 1 along $x$,
  particle 2 along $x$, and particle 3 along $x$ is {\em not}
  $+1$, but rather $-1$.
Taken together, the four spectra in Fig.\,1 clearly demonstrate the GHZ
  correlations predicted by quantum mechanics.

Although the NMR experiment demonstrates GHZ correlations, it cannot eliminate
  local interactions as the cause of those correlations because of the
  close proximity of the nuclei involved.
However, no known interaction in the natural Hamiltonian
  could be responsible for the ``communication''
  between the nuclei that caused the GHZ correlations.
Specifically, measurement step (D) required only $4.043$\,ms to implement;
  this represents the amount of time {\em after}\/ the GHZ preparation steps
  (A)$\sim$(C)---common to all four experiments---but before data acquisition,
  in which the culpable classical local interaction could ``inform'' the spins
  of which axes were being measured for which spins and then adjust the phase
  of the acquisition signal to reveal the quantum mechanical correlations.
Since the fastest coupling term in the natural Hamiltonian is slower than
  $4.043$\,ms ($1/2J_{12}=9.36$\,ms), the culpable ``classical demon,'' if it
  exists, would have to be a non-standard interaction.

In addition, the NMR experiment cannot completely rule out classical hidden
  variables.
The effective pure state used to demonstrate GHZ correlations is in fact a
  mixed state of an ensemble in which individual members are unitarily
  transformed from a thermal state.
It can be shown that such states can be described by non-negative discrete
  Wigner functions\cite{l:dwf,c:hv}.
Accordingly, the results of the experiment could be ``explained'' by a
  hypothetical ensemble of classical systems each of which possesses a
  definite state for its hidden variables.
As in the case of local interactions in the previous paragraph, however, to be
  consistent with the measured properties of the states (3-5) above the
  preparation of such an ensemble from the thermal ensemble with which the
  experiment begins would require a conspiracy of non-standard interactions. 

In conclusion, this paper reported on an experiment that displays GHZ
  correlations using nuclear magnetic resonance.
Although the NMR experiment performed cannot entirely rule out classical local
  hidden variables, it nonetheless provides an unambiguous and statistically
  significant demonstration of the apparently paradoxical GHZ correlations.
In addition, the experimental demonstration of GHZ correlations using NMR
  underscores the ability of NMR to perform significant experiments
  in fundamental quantum mechanics.

\vfil
\noindent{\it Acknowledgements:} This work was supported by DARPA.
\vfil\eject

\vfill\eject

\begin{thebibliography}{1}
\bibitem{epr:cqmdrcc}A. Einstein, B. Podolsky, and N. Rosen, Can quantum-
  mechanical description of reality be considered complete?,
  {\it Phys.\ Rev.\ }47:777-780 (1935).
\bibitem{b:qt} D. Bohm, {\it Quantum Theory}, Prentice-Hall: Englewood Cliffs
  (1951).
\bibitem{b:asiqtthv} D. Bohm, A suggested interpretation of the quantum
  theory in terms of ``hidden" variables, I and II, {\it Phys. Rev.},
  85:166-179 (1952).
\bibitem{b:oeprp} J. Bell, On the Einstein Podolsky Rosen paradox,
  {\it Physics} 1:195-200 (1964).
\bibitem{agr:etrltvbt} A. Aspect, P. Grangier, and G. Roger, Experimental
  tests of realistic local theories via Bell's theorem,
  {\it Phys. Rev. Lett.}, 47:460-463 (1981).
\bibitem{agr:ereprbg} A. Aspect, P. Grangier, and G. Roger, Experimental
  realization of Einstein-Podolsky-Rosen-Bohm gedankenexperiment: a new
  violation of Bell's inequalities, {\it Phys. Rev. Lett.}, 49:91-94 (1982).
\bibitem{ghsz:btwi} D. Greenberger, M. Horne, A. Shimony, and A. Zeilinger,
  Bell's theorem without inequalities, {\em Am. J. Phys.}, 58:1131-1143 (1990).
\bibitem{gc:bsrqc} N. Gershenfeld and I. Chuang, Bulk spin-resonance quantum
  computation, {\em Science}, 275:350-356 (1997).
\bibitem{cfh:eqcnmrs} D. Cory, A. Fahmi, and T. Havel, Ensemble quantum
  computing by nuclear magnetic resonance spectroscopy, {\em Proc. Natl.
  Acad. Sci.} 94:1634-1639, 1997.
\bibitem{cph:nmrsaeapqc} D. Cory, M. Price, and T. Havel, Nuclear magnetic
  resonance: an experimentally accessible paradigm for quantum computing,
  {\it Physica D}, 120:82-101 (1998).
\bibitem{l:maghze} S. Lloyd, Microscopic analogs of the Greengerger-Horne-
  Zeilinger experiment, {\it Phys. Rev. A}, 57:R1473-R1476 (1998).
\bibitem{lkzcm:nmrghz} R. Laflamme, E. Knill, W. Zurek, P. Catasti, and S.
  Mariappan, NMR GHZ, {\it Phil. Trans. Roy. Soc. Lond.}, A356:1941-1948
  (1998).
\bibitem{m:qmr} N. Mermin, Quantum mysteries revisited, {\it Am. J. Phys.},
  58:731-734 (1990).
\bibitem{bdsw:mseqec}C. Bennett, D. DiVincenzo, J. Smolin, and W. Wootters,
  Mixed-state entanglement and quantum error correction, {\it Phys.\ Rev.\ A}
  54:3824-3851 (1996).
\bibitem{sm:de}J. Schlienz and G. Mahler, Description of entanglement,
  {\it Phys.\ Rev.\ A} 52:4396-4404 (1995).
\bibitem{l:dwf} S. Lloyd, Discrete Wigner Functions, in preparation.
\bibitem{c:hv} S.L. Braunstein, {\it et al.}, Separability of
  very noisy mixed states and implications for NMR quantum computing,
  {\it quant-ph}/9811018 (1998). 

\end{thebibliography}
\end{document}